\documentstyle[12pt]{article}
\def\bold#1{\setbox0=\hbox{$#1$}%
      \kern-.025em\copy0\kern-\wd0
      \kern.05em\copy0\kern-\wd0
      \kern-.025em\raise.0433em\box0 }
\newcommand{\be}{\begin{eqnarray}}
\newcommand{\ee}{\end{eqnarray}}
\newcommand{\bearr}{\begin{array}}
\newcommand{\eearr}{\end{array}}
\newcommand{\nsp}{\noindent}

\def\pmu{\partial_{\mu}}

\def\pnu{\partial_{\nu}}
\def\vmu{V_{\mu}}
\def\vnu{V_{\nu}}

\def\Dmup{{\cal D}_{\mu}}

\def\omu{\omega_{\mu}}

\def\rmu{\rho_{\mu}}

\def\phimu{\phi_{\mu}}

\def\ksmu{K^*_{\mu}}

\def\gasb{\Gamma_{sb}}
\def\gano{\Gamma_{an}}
\def\gawz{\Gamma_{WZ}^0}

\def\mpi2{m_{\pi}^2}
\def\meta2{m_{\eta}^2}
\def\mk2{m_K^2}
\def\mks{m_{K^*}}
\def\zks{Z_{K^*}}

\def\xil{\xi_L}
\def\xir{\xi_R}
\def\xilr{\xi_{L\left( R \right)} }
\def\xild{\xi_L^\dagger}
\def\xird{\xi_R^\dagger}
\def\xilrd{\xi_{L\left( R \right)} ^\dagger}

\def\fpi{f_{\pi}}
\def\fpic{f_{\pi}^2}

\def\dmul{{\Dmup \xil \ \xild }}
\def\dmur{{\Dmup \xir \ \xird }}

\def\al{a_L}
\def\ar{a_R}
\def\alr{a_{L\left( R \right)} }

\def\Upi{U_{\pi}}
\def\Upid{U_{\pi}^{\dagger}}

\def\Upih{\sqrt{U_{\pi}} }
\def\Ukh{\sqrt{U_K}}

\def\Tavec{\vec{\tau}}
\def\Pivec{\vec{\pi}}

\def\kad{K^\dagger}

\def\lsu2{{\cal L}_{SU\left( 2 \right)}}

\def\vlmu{v_{\mu}}

\def\qlmu{q_{\mu}}
\def\aumu{a^{\mu}}
\def\almu{a_{\mu}}
\def\Dumuk{D^{\mu}K}

\def\Dumukd{\left( D^{\mu}K \right) ^\dagger}
\def\Dlmukd{\left( D_{\mu}K \right) ^\dagger}

\def\vqlmu{{\left( \vlmu + \qlmu \right) }}

\def\Dlnukd{\left( D_{\nu}K \right) ^\dagger}

\def\sfr2{{sin^2 F \over r^2}}

\def\fp2sf2{ \left( F'^2 + 2 \ \sfr2 \right)}

\begin{document}
\begin{titlepage}
\begin{center}
\hfill TAN-FNT-93-17

\setcounter{footnote}{1}
\def\thefootnote{\fnsymbol{footnote}}
\vspace*{2.0cm}
{\large\bf HYPERONS IN THE BOUND STATE APPROACH WITH VECTOR MESONS}
\vskip 1.5cm
{\large CARLOS L. SCHAT and NORBERTO N. SCOCCOLA
\footnote{
Fellow of the Consejo Nacional de Investigaciones Cient{\'{\i}}ficas
y T\'ecnicas.}}
\vskip .2cm
{\large 
Physics Department, Comisi\'on Nacional de Energ\'{\i}a At\'omica,}\\
{\large  Av.Libertador 8250, (1429) Buenos Aires, Argentina}
\vskip 2.cm
December 30, 1993
\vskip 2.cm 
{\bf ABSTRACT}\\
\begin{quotation}
We investigate a model for hyperons based on the bound state approach
in which vector mesons are explicitly incorporated. We show that for 
empirical values of the mesonic parameters the strange hyperon spectrum 
is well reproduced. We also discuss the extension of the model to
heavier flavors. We show that the explicit presence of the
heavy vectors leads to good predictions for the heavy baryon masses.
\end{quotation}
\end{center}
\end{titlepage} 

\setcounter{footnote}{0}
\def\thefootnote{\arabic{footnote}}

The bound state approach \cite{CK85} has become a very useful tool for
the study of hyperon properties \cite{CKVAR}. In this approach
strange hyperons are described as soliton-kaon bound systems. To find the
soliton-kaon interactions one usually starts with an effective
chiral lagrangian supplemented with proper chiral symmetry breaking
terms. Up to now, most of the calculations have been done using 
effective lagrangians where only pseudoscalar meson fields were 
included. There are several reasons to believe, however, that these are not 
the only degrees of freedom to be taken into account. Within the chiral
soliton model vector
mesons are known to improve the description of the
nucleon properties \cite{MEI88}. 
Moreover, it has been recently argued \cite{RRS90,WJM93} that 
they are of fundamental importance in the
extension of the bound state description to heavier flavors. 
It is now well-known that in the heavy quark limit of QCD
the heavy vector meson (i.e. $B^*$) becomes degenerate with the
heavy pseudoscalar (i.e. $B$) and therefore should be explicitly included 
in the effective action \cite{IW91}. 
In the past, some attempts have been
done to explicitly incorporate vector mesons in the bound state
model. In Ref.\cite{SNNR88} a model with an explicit $\omega$ vector
meson was studied. In Ref.\cite{SMNR89} although the full nonet of 
vector mesons was included, the strange vector meson ($K^*$) was
integrated out before solving the equations of motion. In the 
present paper we will deal with the full vector meson nonet
throughout. In addition, a more general  chiral symmetry breaking action 
that includes derivative-type terms will be used. 
Finally, although most of this work will
be devoted to the description of low-lying positive parity strange
hyperons some results for charmed and bottom baryons will also be
given.
    
To incorporate vector mesons in the chiral lagrangian we use the
hidden gauge approach\,\cite{BKY88}. 
This scheme is based on the identification
of the ``strong" vector mesons as gauge bosons of the hidden gauge symmetry
of the non-linear sigma model. The corresponding effective action is given
by 
\be
\Gamma &=& \Gamma_0+\Gamma_{sb}+\Gamma_{an},
\label{gamma}
\ee
\nsp
where $\Gamma_0$ is the non-anomalous chirally symmetric contribution
to the action

\be
\Gamma_0 &=&
\int d^4x\left\{\frac{1}{2g^2}\  Tr\ F_{\mu\nu} F^{\mu\nu} \right.\nonumber\\
& & -\left.{\fpic \over4} \left[ Tr \left({\dmul} -
{\dmur} \right)^2 + a\ Tr \left({\dmul} + {\dmur} \right)^2 \right]\right\}.
\label{gana}
\ee

The covariant derivatives are
\be
\Dmup \xilr = \left( \pmu - \vmu \right) \xilr
\label{cd} 
\ee

\vspace{2cm}

\nsp
and the field strength tensor is
\be
F_{\mu\nu} = \pmu \vnu - \pnu \vmu - \left[ \vmu , \vnu \right].
\label{fs}
\ee
\nsp
The antihermitian vector meson field $\vmu$ is defined as
\be
V_\mu=
{1\over2}
\left(\begin{array}{cc}\omega_\mu + \rho_\mu &\sqrt2{K_\mu}^* \\
-\sqrt2 {{K_\mu}^*}^\dagger& \sqrt2 \phi_\mu\end{array}\right)
\ee
\nsp
where $\omu$ corresponds to the isoscalar vector meson, $\rmu$ to the
isovector vector meson, $\ksmu$ to the isodoublet of strange vector
mesons and $\phimu$ to the isosinglet strange vector meson.
Although the constant $a$ in Eq.(\ref{gana}) is, strictly speaking, an 
arbitrary constant within the framework adopted here, the value $a=2$ 
leads to a very successful meson phenomenology.
In fact, using that value one obtains the KSRF relation as well as
vector meson dominance\cite{BKY88}. 
In what follows, we will take the value $a=2$ throughout.

In Eq.(\ref{gamma}) we have included a chiral symmetry breaking 
term $\gasb$ whose expression is given by \cite{SSW93}

\be
\gasb &=& \int d^4 x \Bigg\{ {m_\pi^2 \fpic + 2 \mk2 f_K^2 \over{12}} \ 
Tr\left[ U - 1 \right] 
+ {m_\pi^2 \fpic - m_K^2 f_K^2 \over{6}} \ 
Tr\left[ \sqrt{3} \lambda_8 \ U  \right] 
\nonumber \\
& & - Tr\left[ (1-\sqrt{3} \lambda_8) 
\left( \alpha\ ({\cal D}_\mu \xi_L)^\dagger {\cal D}^\mu \xi_R
+ \beta\ U^\dagger \partial_\mu U \partial^\mu U^\dagger +
\gamma\ \xi_L^\dagger F_{\mu\nu} F^{\mu\nu} \xi_R 
\right) \right] \Bigg\}
\nonumber \\
& & + h.c.
\label{gasb} 
\ee

\nsp
Here, we have considered only those chiral symmetry 
breaking terms in Ref.\cite{SSW93} 
which are linear in the quark mass matrix and satisfy the 
OZI rule. In addition we have neglected the
non-strange quark masses in the derivative-type symmetry breakers. 
These approximations do not affect in any
significant way the properties of the mesons we are interested in.
In  Eq.(\ref{gasb}), $U$ is the chiral field ($=\xi_L^\dagger \xi_R$),
$m_{\pi}$ is the pion mass, 
$m_K$ the kaon mass and $f_\pi$
and $f_K$ are the pion and kaon decay constants respectively.
The coefficients $\alpha$, $\beta$ and $\gamma$ are defined
as follows
\be
\alpha & = & {1\over{3 g^2}} \left( m_v^2 - \zks^2 \mks^2 \right) \\
\beta  & = & {1\over{12}} \left( \fpic - f_K^2 + 
3\ \alpha \right) \\
\gamma & = & {1\over{6g^2}} \left( 1 - \zks^2 \right)
\ee 
\nsp
where $m_v$ ($= \sqrt{a} g f_\pi$) 
is the non-strange vector meson mass, $\mks$ is the
$K^*$ mass and $\zks$ is the renormalization constant of the $K^*$
meson field. 
 
Finally, the anomalous action $\gano$ is given by
\be
\gano = \gawz + {i N_c\over{24\pi^2}}
 \int \left( Tr \left( \al^3 \ar - \ar^3 \al \right)
 - Tr \left( \al \ar \al \ar \right)
 \right) 
\label{gano}
\ee
\nsp
where $\gawz$ is the irreducible Wess-Zumino action containing only the
pseudoscalar fields and $ \alr = \Dmup \xilr \ \xilrd dx^{\mu} $.
Eq.(\ref{gano}) is a particular combination of the general lowest
order anomalous action given in Refs.\cite{FKTUY85,JJMPS88}. 
As done in Ref.\cite{SMNR89} we will simplify its rather cumbersome 
form by expressing the anomalous $\rho$ and $K^*$ vector meson couplings 
in terms of the pseudoscalar fields.
With this approximation our model turns out to be an $SU(3)$ 
generalization of the minimal model studied in Ref.\cite{MKW87}.
No major difference in the predictions of the baryon observables has
been found in models where the full anomalous action was
included \cite{JJMPS88}. 

To study the baryon sector of the model 
we have to define an ansatz for the $\xil$ and $\xir$ fields.
As in Ref.\cite{SMNR89} we use 
\be
\xild &=& \Upih \Ukh\nonumber\\
\xir &=& \Ukh \Upih  \label{2.9}
\ee
where

\be
\Upi=\left(\begin{array}{cc}N^2&0\\\ 0&1\end{array}\right),
\quad \quad 
N = \exp\left[ i {{{\Tavec} \cdot {\Pivec} } \over{2\fpi} }\right],
\ee 
\be
U_K= \exp\left[{i\sqrt2 \over{\fpi}} 
\left( \begin{array}{cc} 0&K\\\ K^\dagger&0 \end{array} \right)
\right] , 
\qquad \qquad 
K=\left(\begin{array}{c}K^+\\K^0\end{array}\right).\nonumber
\ee

\nsp
Following the usual steps in the bound state approach we can now
obtain the soliton-kaon action by inserting the ansatz
Eq.(\ref{2.9}) into  the effective action and expanding up to second
order  in kaon fields. To make this expansion consistently, only terms up to
second order in $K^*$ should be retained. The resulting lagrangian density
can be separated in two classes of terms.
The first ones reduce to the lagrangian of interacting nonstrange fields 
(pions and $\omega$ and $\rho$ vector mesons) which we will denote $\lsu2$.
The minimization of the corresponding action using the Skyrme-Wu-Yang
ansatz provides the $\pi$, $\omega$ and $\rho$ solitonic 
fields. Details of this procedure as
well as the explicit expression of $\lsu2$ can be found in 
Ref.\cite{MKW87}. The rest of the terms describe the 
coupled system of quantum $K$ and $K^*$ fields moving in
the presence of the background fields. It is interesting to note that
to quadratic order in $K$ fields, the $\phi$ vector meson is completely
decoupled from the $KK^*$ system. Therefore its contribution to
$\cal L$ will be ignored. In terms of the
renormalized $K$ and $K^*$ fields\footnote{
To obtain the canonical form of the free action in the limit in which 
meson-soliton interactions are neglected the strange meson fields have 
to be renormalized. This is due to the presence of the 
derivative-type symmetry breakers in Eq.(\ref{gasb}).} 
the resulting meson-soliton 
lagrangian reads 
 
\be
{\cal L} & = & \lsu2 + \Dlmukd \Dumuk + \kad \almu \aumu K 
- m_K^2 \kad K 
\nonumber \\
& & + {1\over{2 g^2 f_K^2}} (3m_v^2 - 2 \zks^2 \mks^2) 
\left[ \kad \vqlmu \Dumuk - \Dumukd \vqlmu K \right]  
\nonumber \\
& & - {1\over{g^2 f_K^2}} \kad \Bigg[ (2 m_v^2 - \zks^2 \mks^2)
a_\mu a^\mu + (m_v^2 - \zks^2 \mks^2) (q_\mu + v_\mu)  
(q^\mu + v^\mu) 
\nonumber \\
& & \qquad \qquad \qquad
- { 1 - \zks^2 \over2} q_{\mu\nu} q^{\mu\nu}
+ {m_\pi^2 g^2 \fpic \over{4}} (\Upi + \Upid - 2) \Bigg] K
\nonumber \\
& & +i {m_v^2 \over{g^2 f_K \zks}} \left[ \kad a_\mu K^{*\mu}
+ K_\mu^{*^\dagger} a^\mu K \right]
\nonumber \\
& & - {1\over{2g^2}} \left[ K^{*^\dagger}_{\mu\nu} K^{*\mu\nu} 
- 2 \mks^2 K_\mu^{*^\dagger} K^{*\mu} +
{2\over{\zks^2}} K^{*^\dagger}_\mu q^{\mu\nu} K^*_\nu \right] 
\nonumber \\
& & + {i N_c \over{ 8 f_K^2}} \Bigg[ B_\mu 
\left( \Dumukd K - \kad \Dumuk \right)
\nonumber \\
& & \qquad \qquad + 3 \omu B^\mu \kad K +
{4\over{3\pi^2}} \epsilon^{\mu\nu\beta\eta} \omu \Dlnukd a_\beta 
D_\eta K \Bigg]
\ee

\nsp
where

\be
D_\mu K & = & \partial_\mu K + v_\mu K , \nonumber \\
q_\mu & = & {1\over2} (\rho_\mu + \omu) , \nonumber \\
q_{\mu\nu} & = & \partial_\mu q_\nu - \partial_\nu q_\mu -
\left[ q_\mu , q_\nu \right] , \nonumber \\
\left( \begin{array}{c}v_\mu \\ a_\mu \end{array} \right ) 
& = & {1\over2} 
\left( N^\dagger \partial_\mu N \pm N \partial_\mu N^\dagger \right),
\nonumber \\
B^\mu & = & {1\over{24 \pi^2}} \epsilon^{\mu\nu\beta\eta}
Tr \left[ \Upid ( \partial_\nu \Upi ) \Upid (\partial_\beta \Upi)
\Upid (\partial_\eta \Upi) \right],
\nonumber \\
K^*_{\mu\nu} & = & \partial_\mu K^*_\nu - \partial_\nu K^*_\mu -
q_\mu K^*_\nu + q_\nu K^*_\mu.
\ee   

To find the ${\cal O}(N_c^0)$ corrections to the hyperon properties 
we should now obtain 
the equations of motion of the $K$ and $K^*$ meson fields in the 
presence of the static soliton background. Since in this work we 
are only interested in the properties of the low-lying positive
parity hyperons it is enough to find the ground state of the 
coupled meson system. For that particular state the meson fields
can be expressed as follows
\be
K(\vec r, t) & = & {1\over{\sqrt{4\pi}}} \ k(r,t) 
\ \vec \tau \cdot \hat r \ \chi \nonumber
\\
K^*_0(\vec r,t) & = & {g\over{\sqrt{4\pi}}} \ t_0(r,t) \ \chi
\\
\vec K^*(\vec r,t) & = & {g\over{\sqrt{4\pi}}} \left[ t_2(r,t) \ \hat r - 
         i \ t_3(r,t) \ \hat r \times \vec \tau \right] \chi \nonumber
\ee
where $\chi$ is the two-component isospinor.
It is interesting to note that in the ground state the
$K^*$ vector meson field can be expressed in terms of only 3
time-dependent radial functions.
In addition, as usual when dealing with
vector fields, there is no dynamics associated with $t_0$. As a
consequence, the diagonalization of the corresponding
meson hamiltonian leads, for the ground state, to a system 
of 3 coupled eigenvalue equations. Their lengthy expressions
will be given elsewhere. The numerical solution of this 
eigenvalue problem provides 
the ${\cal O}(N_c^0)$ correction to the 
hyperon masses $\epsilon$ together with the meson eigenfunctions. 

Finally, the splittings between hyperons with the same value
of strangeness but different spin and/or isospin are obtained
to ${\cal O} (N_c^{-1})$ by introducing the soliton $SU(2)$
collective rotations $A(t)$ 
\be
U_\pi &\rightarrow& A U_\pi A^\dagger 
\nonumber \\
K &\rightarrow& A K \\
K^*_\mu &\rightarrow& A K^*_\mu.
\nonumber
\ee
As it has been shown in Ref.\cite{MKW87} to this order some 
new components of the $\rho$ and $\omega$ vector fields are excited. 
They contribute, together with the rest of the background fields, to 
the $SU(2)$ moment of inertia $\Theta$. Details of the quantization
of the $SU(2)$ sector of the present model can be found in 
Ref.\cite{MKW87}. The strange sector contributes to the
rotational lagrangian with a term proportional to the scalar 
product of the induced meson spin times the soliton angular 
velocity \cite{CK85}. The constant of proportionality is the
so-called hyperfine splitting constant $c$. It can be calculated
as the volume integral of some complicated function
of the $SU(2)$ fields and the $K$ and $K^*$ eigenfunctions. 
Its explicit form will be also given elsewhere. 
    
Summing up all the contributions to the meson-soliton hamiltonian 
and taking matrix elements between hyperon states one obtains the 
bound-state mass formula. It reads\cite{CK85,CKVAR} 
{\small
\be
M_{I,J,S} & = & M_{sol} + |S| \epsilon +
{1\over{2\Theta}} \left[ c J(J+1) + (1-c) I(I+1) + 
{c (c-1)\over4} |S| (|S| + 2) \right]
\label{massf}
\ee }
where $I$, $J$ and $S$ are the hyperon isospin, spin and
strangeness, respectively.
     
In our numerical calculations we set all the parameters
that appear in the effective lagrangian to their empirical values.
Namely, for the parameters that determine the hadron properties
in the $SU(2)$ sector we use $f_\pi = 93 \ MeV$, $m_\pi = 138 \ MeV$,
$g = 5.85$ and $m_v = m_\rho = m_\omega = 770 \ MeV$. Using these
values the soliton mass turns out to be $M_{sol} = 1470 \ MeV$
while the $SU(2)$ moment of inertia is $\Theta = 0.82 \ fm$. Although
the calculated moment of inertia compares fairly well with the
best fit value $\Theta^{BF} = 1.01 \ fm$ needed to reproduce
the empirical $\Delta-N$ mass splitting, the soliton mass is much
larger than $M_{sol}^{BF} = 866 \ MeV$ needed to fit
the nucleon mass. 
This large value of $M_{sol}$ 
obtained for empirical $f_\pi$ is an old problem in
skyrmion physics. Quite recently it has been argued
\cite{MK91,HOL92} that this problem can be solved 
by taking into account the ${\cal O} (N_c^0)$ Casimir
corrections due to pion loops. This might justify the 
traditional attitude we follow here of ignoring this
problem by considering only hyperon masses taken with respect 
to the nucleon mass. For the mesonic parameters in the $SU(3)$ sector 
of the model we use
$f_K = 114 \ MeV$, $m_K = 495 \ MeV$ and $\mks = 892 \ MeV$.
Within the present model, however, the value of $\zks$ cannot
be consistently determined from the strange meson properties\cite{SSW93}. 
In fact, while the $K^* \rightarrow K\pi$ decay ratio favors the value 
$\zks$ = 0.75 from the empirical $\phi$ meson mass one obtains 
$\zks \simeq 0.96$. For this reason we will consider $\zks$ as a free
parameter within that range of values and study the behavior of the
bound meson energy $\epsilon$ and the hyperfine splitting constant
$c$ as a function of it. Results are shown in Fig.1. We observe that 
$\epsilon$  depends quite strongly on $\zks$. On the other 
hand, the constant $c$
is less sensitive to $\zks$. In the context of the
Skyrme model a similar behavior has been found as a function of
the kaon renormalization constant \cite{RS91}. We also see that
for $\zks = 0.85$ the calculated values of $\epsilon = 217 \ MeV$
and $c=0.65$ are very close to the best fit values\cite{CKVAR} 
$\epsilon^{BF}= 218 \ MeV$ and $c^{BF} = 0.66$. In addition, 
this value of $\zks$ agrees very well with $\zks = 0.84$ used in 
Ref.\cite{SSW93} to obtain a good overall fit of 
the meson properties 
\footnote{For $\zks=0.85$ we obtain
$\Gamma(K^* \rightarrow K\pi) = 39 \ MeV$ and $m_\phi=1.12 \ GeV$
to be compared with the empirical values $50 \ MeV$ and $1.02 \ GeV$
respectively.}. 

Before presenting our predictions for the hyperon
masses some remarks are in order. First, we should stress the 
importance of the derivative-type symmetry breaking terms. Namely,
if we set $\alpha = \beta = \gamma = 0$ in Eq.(\ref{gasb})
we obtain $\epsilon = 110 \ MeV$. As in the case of the Skyrme 
model strange mesons are overbound in the
absence of such terms \cite{RS91}. Second, when comparing this value
with the results of Ref.\cite{SMNR89} where these terms have
been neglected and strange vector mesons have been completely 
integrated out
we observe that the explicit presence of these mesons tends
to increase the binding energy. In the present calculation
this effect is compensated by the derivative-type symmetry breakers.

In Table 1 we show our prediction for the strange hyperon
mass spectrum. We observe that the model predicts the mass
differences within few percent. The main source of disagreement
is the somewhat small value of the calculated $SU(2)$ moment of inertia.
It is interesting to note that if this moment of inertia is
taken to its best fit value
all the hyperon masses are predicted within roughly $10 \ MeV$
from their empirical values.  
It would be interesting to see whether effects like i.e. pion loop 
corrections could contribute to increase the value of the
predicted $SU(2)$ moment of inertia.

Finally, we will briefly discuss the extension of the present model 
to the charm and bottom sectors. As already mentioned 
in those cases heavy vector mesons are expected to play a crucial 
role in the determination of heavy meson binding energies.
In the charm sector, using $f_D = 158 \ MeV$ and $Z_{D^*}= 0.47$, we 
predict $\Lambda_C~-~N~=~1343 \ MeV $ and $\Sigma_C - N = 1553 \ MeV$ 
that compare very well with the empirical values $ 1346 \ MeV$ and 
$1514 \ MeV$ respectively. In the bottom sector, using $f_B = 149 \
MeV$  and $Z_{B^*} = 0.32$ we predict  $\Lambda_B - N = 4704 \ MeV$ 
and $\Sigma_B^* - \Sigma_B \simeq 8 \ MeV$.
The $\Lambda_B - N$ splitting agrees very nicely with the
empirical value $4702 \ MeV$ while the hyperfine splitting 
$\Sigma_B^* - \Sigma_B$ is very close to zero as required by heavy
quark symmetry. In all these calculations the heavy meson masses
have been set to the empirical values
given in Ref.\cite{PDT92}. From the results above it is clear that the 
overbinding problems found in previous calculations where only pseudoscalars 
were present
\cite{RRS90} can be completely overcome in the present model
without affecting the predictions for the hyperfine splittings. 
A very interesting point for further investigation is the connection
between the present model and those based on explicit heavy quark symmetry
\cite{WJM93,KOR93,SCH93}.

In conclusion, we have studied a model for hyperons based on the bound
state approach in which 
vector mesons are explicitly incorporated both in the light
and massive sectors. We have shown that good predictions
for the hyperon spectrum can be obtained for empirical
values of the mesonic parameters. Derivative-type chiral symmetry
breaking terms are crucial in obtaining these results.
Finally, we have also shown that when extended to the charm 
and bottom sectors the present model eliminates the overbinding 
found in models where only pseudoscalars are present.

Discussions with E. Jenkins, A. Manohar, Y. Oh, D.O. Riska, M. Rho and
K. Yamawaki are greatly acknowledged. 
    
\vspace{1.cm}
\noindent
 
\pagebreak

 
\vspace*{2.5cm}
\noindent
{\Large {\bf Figure and Table Captions}}

\vspace{1.5cm}

\noindent
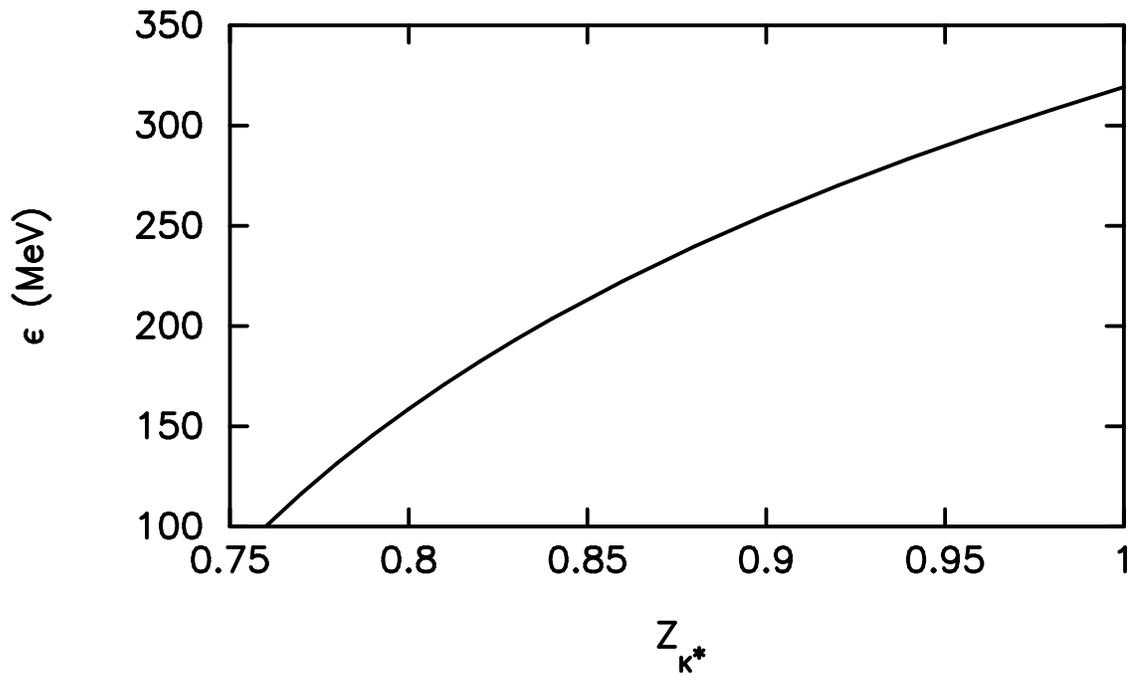
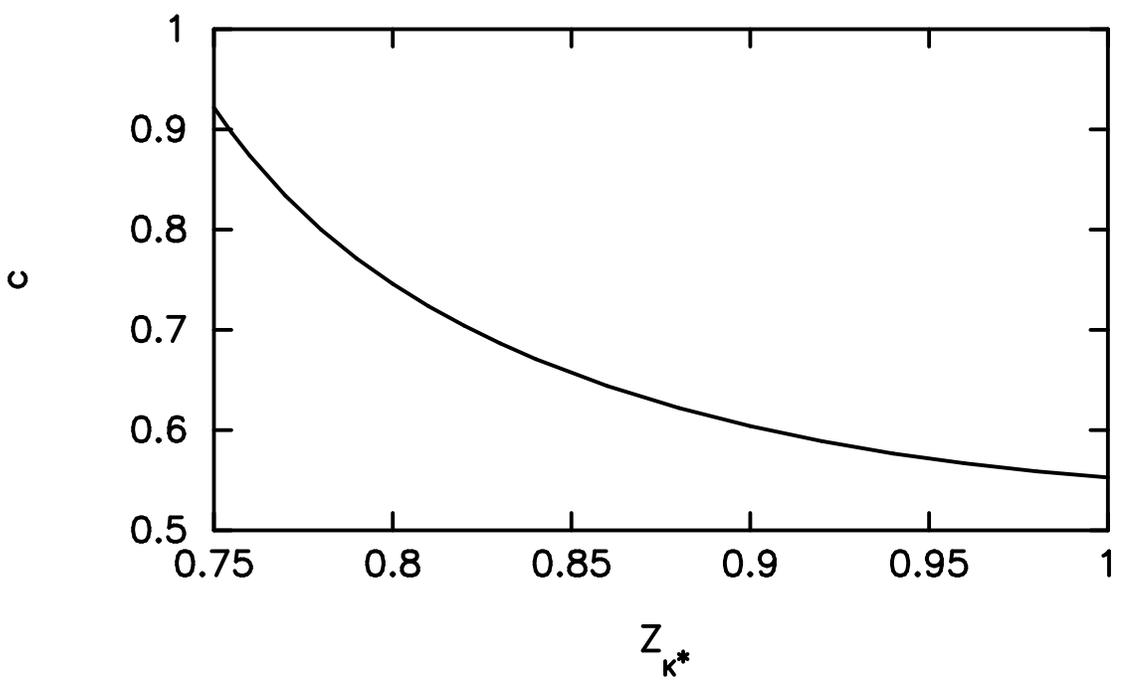
{\bf Fig. 1}: Meson eigenenergy $\epsilon$ and hyperfine
splitting constant $c$ as functions of the $K^*$ renormalization
constant $Z_{K^*}$. 

\vspace{1cm}
\noindent
{\bf Table 1}:  Strange hyperon masses  (in $MeV$)
taken with respect to the calculated nucleon mass.
The experimental values are given in the first column. They
correspond to the mass differences between
isopin multiplet averages, with uncertainties given by half the sum
of the difference of extremum values within each multiplet. The 
calculated values are given in the second column. The mesonic
parameters used in this calculation are $f_\pi = 93 \ MeV$, 
$m_\pi= 138 \ MeV$,
$f_K=114 \ MeV$, $m_K = 495 \ MeV$, $g=5.85$, $m_v=770 \ MeV$,
$m_{K^*} = 892 \ MeV$ and $Z_{K^*} = 0.85$. 

\pagebreak
\vspace*{3cm}
\centerline{ \Large {\bf Table 1} }
\vspace{1.5cm}
\begin{center}
 \begin{tabular}{|c|c|c|} \hline
\hspace{.2cm} Particle \hspace{.2cm} & 
\hspace{1cm} Exp. \hspace{1cm} & 
\hspace{.4cm} This model \hspace{.4cm} \\
 \hline
 $\Lambda$                & $ 177 \pm 1$ & 165\\
 $\Sigma$                 & $ 254 \pm 5$ & 249\\
 $\Sigma^{*}$             & $ 446 \pm 3$ & 484\\
 $\Xi$                    & $ 379 \pm 4$ & 379\\
 $\Xi^{*}$                & $ 594 \pm 2$ & 614\\
 $\Omega$                 & $ 733 \pm 1$ & 751\\
 \hline \end{tabular}
 \end{center}

\end{document}